\documentclass[%
aip,
apl,
reprint,%
]{revtex4-1}

\usepackage{graphicx}
\usepackage{dcolumn}
\usepackage{bm}

\newcommand*\chem[1]{\ensuremath{\mathrm{#1}}}

\begin{document}

\preprint{AIP/123-QED}

\title{A micro-SQUID with dispersive readout for magnetic scanning microscopy}

\author{F. Foroughi}
\affiliation{University of Grenoble Alpes, N\'eel Institute, Grenoble, France}

\author{J.-M. Mol}%
\email{mol@physik.rwth-aachen.de}
\author{T. Mueller}%
\affiliation{2nd Institute of Physics, Quantum Technology Group, RWTH Aachen University, Germany}%

\author{J. Kirtley}
\author{K.A. Moler}
\affiliation{%
Center for Probing the Nanoscale, Stanford University, California%
}%
\author{H. Bluhm}
\affiliation{2nd Institute of Physics, Quantum Technology Group, RWTH Aachen University, Germany}%
\date{\today}

\begin{abstract}
We have designed and characterized a micro-SQUID with dispersive readout for use in low temperature scanning probe microscopy systems. The design features a capacitively shunted RF SQUID with a tunable resonance frequency from 5 to 12 GHz, micrometer spatial resolution and integrated superconducting coils for local application of magnetic fields. The SQUID is operated as a nonlinear oscillator with a flux- and power-dependent resonance frequency. Measurements for device characterization and noise benchmarking were carried out at 4 K. The measured flux noise above 10 kHz at 4 K is $80\;\mathrm{n}\Phi_0 \mathrm{Hz}^{-1/2}$  at a bandwidth of 200 MHz. Estimations suggest that one can benefit from parametric gain based on inherent nonlinearity of the Josephson junction and reduce the flux noise to $30\;\mathrm{n}\Phi_0\mathrm{Hz}^{-1/2}$ at 100 mK, which corresponds to $7.5\;\mathrm{\mu_B Hz^{-1/2}}$ for a magnetic moment located at the center of the pickup loop.
\end{abstract}

\maketitle
Measuring the magnetic response of mesoscopic samples or mapping it vs. position for extended systems are effective methods for revealing fundamental quantum properties of condensed matter systems. Over the past few decades, many advanced magnetic imaging schemes have been developed, including magnetic force microscopy\cite{MFM}, scanning Hall probe microscopy\cite{Hall}, superconducting interference devices (SQUIDs)\cite{John_review,Hasselbach}, and NV center based magnetometry.\cite{NV} The high sensitivity, low back
action and low power dissipation of SQUIDs makes them attractive for many types of low temperature experiments. One classic example is the central role of scanning SQUID microscopy in tests of pairing symmetry of high-$\mathrm{T_c}$ cuprate superconductors.\cite{John_pairing} Other prominent applications are the thermodynamic characteristics of persistent currents in normal metal rings\cite{Hendrik_persistent} and proof of edge states in topological insulators in the quantum spin Hall regime.\cite{katja}

The magnetic field sensitivity of a SQUID with a given flux-sensitivity increases with its pickup area. However, smaller SQUIDs provide better coupling to smaller samples which leads to higher spin sensitivity\cite{John_review,Huber_SQUID} as the magnetic field of a dipole decreases with $1/r^3$. 
Various approaches have been pursued to reduce the size. The smallest nano-SQUID to date has been fabricated by evaporating Nb or Pb\cite{Zeldov} onto the apex of a sharp quartz tip reaching a loop diameter of below 50 nm and a spin sensitivity of $0.38\;\mathrm{\mu_B}\mathrm{Hz}^{-1/2} $ at a few tens of kHz bandwidth. In another approach, parametric amplification has been harnessed in nano-SQUIDs based on Aluminum junctions\cite{Vijay,Levenson}, which reduces the flux noise down to $30\;\mathrm{n}\Phi_0\mathrm{Hz}^{-1/2} $ with a bandwidth exceeding 60 MHz.\cite{Levenson} Recent works on parametric amplifiers and dispersive readout of superconducting qubits also harness the nonlinearity of the Josephson junction to boost sensitivity.\cite{Castellanos_nature,Sidiqi_review,Sidiqi_HBW} Compared to such size-optimized devices, micro-SQUIDs fabricated using a standard Nb technique have the advantage of on-chip field coils and modulation coils\cite{Huber_SQUID} and can be operated at 4 K allowing for a wide range of applications. However, these come at the cost of larger pick-up loops (a few micrometers) and a resulting poorer spin sensitivity ($200\;\mathrm{\mu_B}\mathrm{Hz}^{-1/2} $). This limitation was previously only overcome by devices with pickup loops defined by focused-ion-beam\cite{koshnik}, but has recently also been achieved by improvements in lithography and shown to offer better spatial resolution as well.\cite{Kirtley2016} 

Here we present a scanning micro-SQUID which is based on the same fabrication technology incorporating another major advancement. Our design exploits the parametric amplification based on nonlinearities of the SQUID to boost sensitivity. The combination of a smaller pickup loop and parametric amplification of the signal leads to better spin sensitivity ($10\;\mathrm{\mu_B}\mathrm{Hz}^{-1/2}$) and higher bandwidth (200 MHz) compared to traditional DC micro-SQUIDs in which the flux sensitivity is mainly limited by internal dissipation.\\

\begin{figure}
\includegraphics[width=8.5cm]{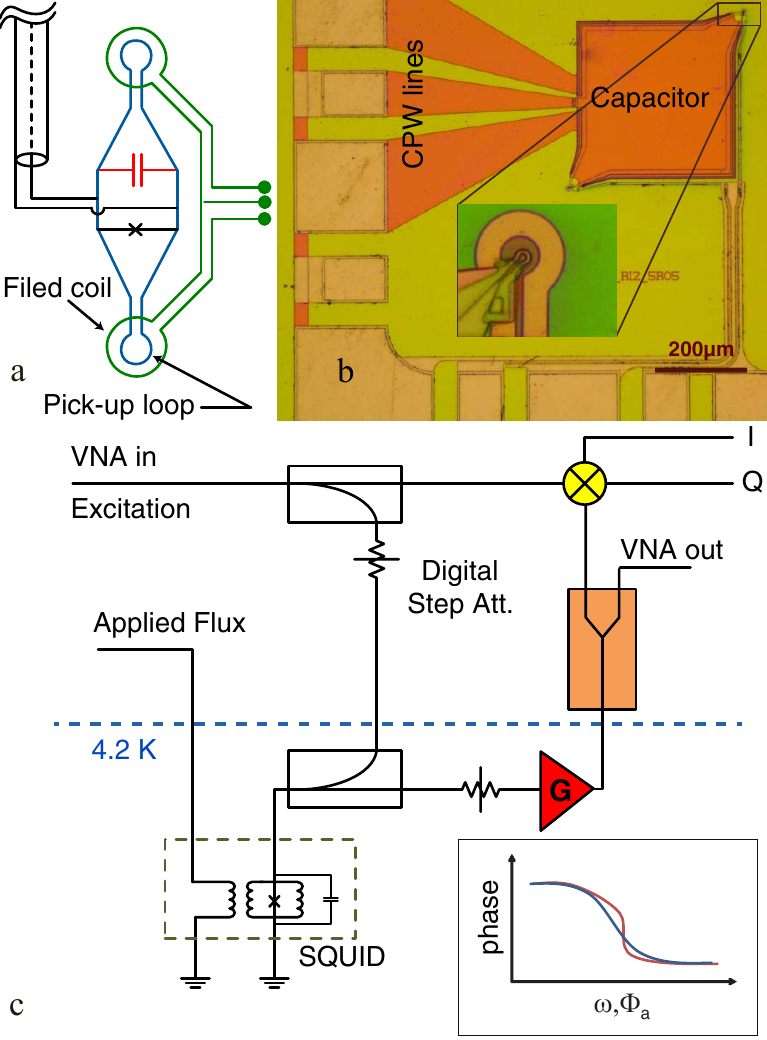}
\caption{\label{fig:optical} (a) Schematic circuit diagram of the SQUID. The geometry is gradiometric and cancels out any homogeneous background magnetic field. Two field coils with a center-tab allow applying magnetic fields to the sample while the total flux threading the SQUID is unchanged. (b) Optical Image of the fabricated chip. (c) Simplified schematic of the read-out setup. The inset shows the effect of increased power (red line) on reflected phase of RF signal as a function of frequency (or applied magnetic flux $\Phi_a$).}
\end{figure}

Our devices were fabricated at \textit{Hypres Inc.} using planarized \chem{Nb}/\chem{AlO_x}/\chem{Nb} trilayer Josephson junction technology, including two planarized \chem{Nb} layers with approximately $0.8\;\mu \mathrm{m}$ minimum feature size. The SQUID consists of a superconducting ring shunted by one Josephson junction with designed critical current of $20\;\mu\mathrm{A}$ [Fig.~\ref{fig:optical}(a)]. With this geometry, the SQUID is gradiometric, i.e. the effect of any homogeneous background magnetic field approximately cancels out. The smallest pickup loops implemented have an inner (outer) diameter of $1\;\mu \mathrm{m}$ ($2\;\mu \mathrm{m}$). The superconducting loop is shunted by an on-chip parallel plate capacitor with amorphous silicon dioxide as dielectric which was designed to have a capacitance of 80 pF. The  superconducting loop and the capacitor form a resonator with a flux dependent resonance frequency from 5-12 GHz. 
The pickup loops are placed on opposite corners of the parallel-plate capacitor at a distance of about $450\;\mu\mathrm{m}$ so that one loop can be located in close proximity to the sample while the other is kept far away from it. External fields can be applied to the sample by local single-turn field coils situated around each pickup loop. The field coils are fully integrated into the chip layout and can also be used to bias the SQUID at its most sensitive point. Compared to using additional modulation coils\cite{Huber_SQUID}, this approach has the advantage of reducing the total device inductance, which leads to a higher sensitivity.\cite{handbook} The effective diameter of each coil is $7.5\;\mu \mathrm{m}$, resulting in a mutual inductance of about $0.34\; \Phi_0 /\mathrm{mA}$ (for the smallest pickup loop). The SQUID and field coils are connected by on-chip $50\; \Omega$ matched coplanar waveguide (CPW) transmission lines to the bonding pads. High frequency signals up to 10 GHz can thus be applied to the SQUID and the sample.\\

As readout, we adopted a dispersive approach.\cite{Vijay} The SQUID is operated as a nonlinear oscillator with a flux- and power-dependent resonance frequency. A continuously applied microwave excitation is reflected by the SQUID with a flux dependent phase shift which can be detected using standard homodyne detection. Fig.~\ref{fig:optical}(c) shows a simplified schematic of this scheme. To thermalize room temperature noise, RF lines are attenuated at low temperature. A directional coupler is used to couple the excitation signal to the SQUID and guide the reflected signal to a semiconductor cryogenic amplifier. The amplifier has a bandwidth of 1-12 GHz so that it does not limit the SQUIDs operation. The measured noise temperature of the amplifier is about 5 K between 3 and 6 GHz. The signal is further amplified by several room temperature amplifiers before being sent to the vector network analyzer (VNA) or to a mixer, where the phase of the reflected signal can be read out. While the I and Q outputs are most convenient for high bandwidth measurements, we characterize the SQUID using a VNA to measure the reflection coefficient ($\Gamma$) of the SQUID as a function of applied magnetic flux and RF power.

\begin{figure}
\includegraphics[width=8.5cm]{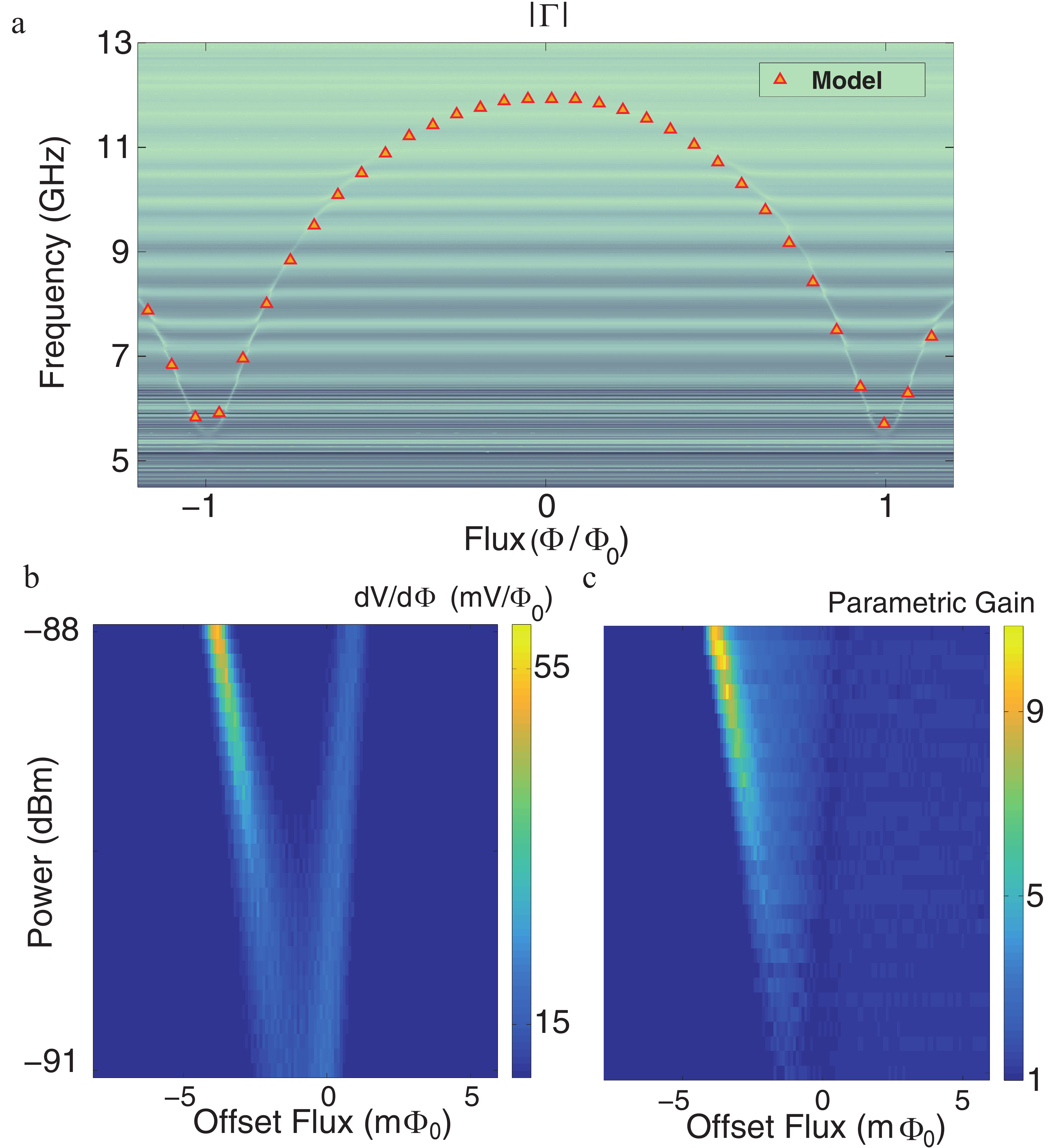}
\caption{\label{fig:raw_data} (a) Dependence of resonance frequency on applied magnetic field and comparison with simulation (triangles). The resonance frequency is a periodic function of the external field with a periodicity of 2$\Phi_0$ (see text). (b) Transduction factor estimation based on excitation power and external bias sweeps. Nonlinearities lead to the splitting into two branches and enhance the transduction factor. (c) Estimated parametric gain $|G|$; only the left branch shows parametric amplification. The data in (b) and (c) were computed by numerical differentiation of the reflected signal. The off-resonant gain was normalized to one to compensate for any microwave signal losses or reflections in the system.}
\end{figure}

An external magnetic flux applied to one pickup loop will induce a circulating current in the SQUID resonator and through the junction which changes the Josephson inductance. Part of the circulating current is shunted through the other pickup loop inductance and reduces the effect of the external flux by a factor of two. Thus, the resonance frequency of the tank circuit and hence the mixer output voltage ($V$) is a periodic function of the applied external magnetic flux ($\Phi_a$) with a periodicity of $2 \Phi_0$ [see Fig.~\ref{fig:raw_data}(a)]. Fitting a simple model of a LC tank-circuit shunted by a Josephson junction, we extracted the critical current of 23 $\mu\mathrm{A}$, a loop inductance of 19 pH and a capacitance of 30 pF. At finite static bias, especially near $\Phi_a\approx \Phi_0$, the resonance frequency is very sensitive to the applied flux, so the highest sensitivity is achieved by biasing the SQUID on resonance at a point with largest slope. 
The transduction factor $dV/d\Phi=G_\mathrm{tot}v_\mathrm{in}\: |\partial\Gamma/\partial\Phi|$ represents the change in the reflected microwave signal at mixer or VNA input in response to the change in the external magnetic flux, where $v_\mathrm{in}$ is the amplitude of the excitation signal applied to the SQUID and $G_\mathrm{tot}$ is total gain of the amplifiers before the mixer and its conversion gain. In the linear regime, $\Gamma$ is independent of the excitation signal and hence the flux sensitivity can be improved by both increasing the excitation amplitude and biasing the SQUID at higher $|d\Gamma/d\Phi|$.

The analysis discussed above assumes that the excitation signal is small enough ($<$ -95 dBm) for the SQUID to behave as a linear inductor. When increasing the power, nonlinear effects arising from the sinusoidal phase-current characteristic of the Josephson junction become important. With moderate excitation, one can harness inherent nonlinearities in the SQUID to enhance the sensitivity [inset of Fig.~\ref{fig:optical}(c)]. However, nonlinearities eventually lead to bifurcation which limits the maximum power level that can be applied to the SQUID. Hence, a crucial factor for optimum noise performance while avoiding bistable operating points is a careful choice of the excitation power and flux bias. When the power is low enough, experimentally we find the lowest-noise operating points near $\Phi_a=0.91 \Phi_0$, which is in the region where the resonance frequency is most sensitive to the applied flux [Fig.~\ref{fig:raw_data}(a)]. To further investigate noise performance of the SQUID, two dimensional sweeps in $\Phi_a$ (small offset flux added to $\Phi_a=0.91 \Phi_0$) and applied power at constant frequency (6.65 GHz) were carried out. Fig.~\ref{fig:raw_data}(b) shows the estimated transduction factor. At low powers there is a single sensitive region at resonance. At higher power this flux sensitive region splits up into two branches; the left branch with enhanced transduction factor and the right one without any enhancement. The distortion of the resonance curve ($\Gamma$) on the left branch leads to a faster change of the reflected phase as a function of the applied flux (increase of  $|d\Gamma/d\Phi|$) and thus improves the transduction factor, as illustrated in the inset of Fig.~\ref{fig:optical}(c).\\

\begin{figure}
\includegraphics[width=8.5cm]{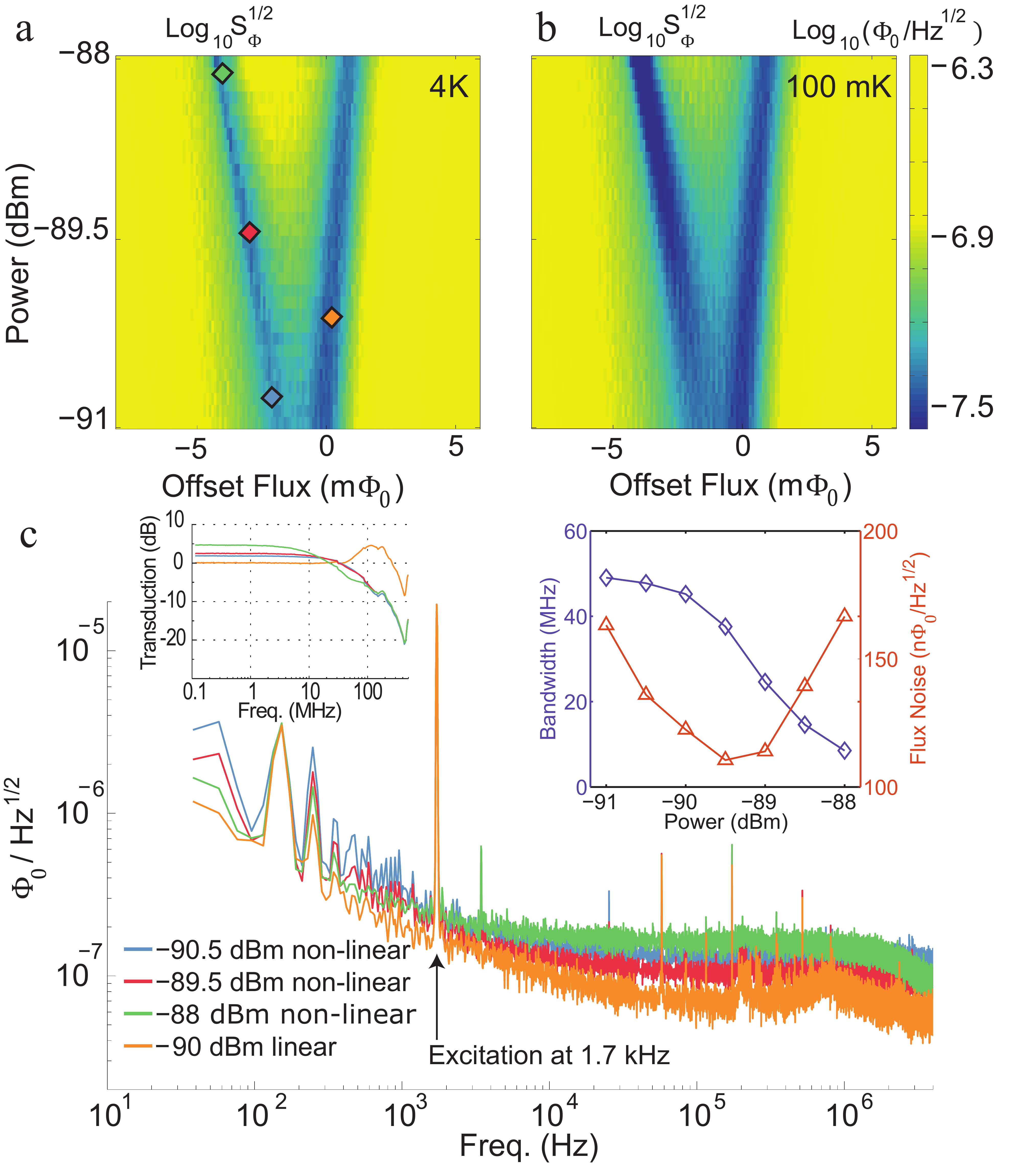}
\caption{\label{fig:estimation} (a, b) Estimated flux noise based on the data from Fig.~\ref{fig:estimation} and the noise model described in the text for $T=4\;\mathrm{K}$, and $T=100\;\mathrm{mK}$ (c) Measured flux noise at 4 K and at different bias points on flux noise estimation. In order to simultaneously measure the transduction coefficient, an ac signal with an amplitude of $20\; \mu\Phi_0$ and frequency of 1.7 kHz was applied to one of the field coils and the voltage noise was obtained by averaging the noise floor between 50 kHz and 100 kHz. The inset shows the dependence of the bandwidth and flux noise on excitation power at the nonlinear branch.}
\end{figure}

While nonlinearities result in an enhancement of the transduction factor, the spectral noise density of the SQUID $S^{1/2}_\Phi={S^{1/2}_V}/{(dV/d\Phi)}$ is also affected by parametric amplification of noise on the excitation line. 
To estimate the total noise level, two contributions have to be taken into account. First, the semiconductor cryogenic amplifier adds its own noise to the reflected signal. Second, noise on the drive line is amplified by the SQUID with parametric voltage gain $G=\Gamma^{-1}(dv_\mathrm{ref}/dv_\mathrm{in})=1+v_\mathrm{in}\Gamma^{-1}(\partial\Gamma / \partial v_\mathrm{in})$, where $v_\mathrm{ref} = \Gamma v_\mathrm{in}$ is the complex voltage amplitude of the reflected signal at the SQUID. The above expression assumes conservation of energy so that changes in $v_\mathrm{in}$ can only affect the phase of $v_\mathrm{ref}$. An estimate of $|G|$ based on the dependence of the measured reflected signal on the excitation power is shown in Fig.~\ref{fig:raw_data}(c). 
This parametric amplification amplifies the power of signals in phase with the excitation by $|G|^2$. The spectral noise density for the sensitive quadrature is thus $S^{1/2}_{V}=\sqrt{k_\mathrm{B}(T_\mathrm{amp}+|G|^2 T_\mathrm{eff})Z_0/2}$, where $k_\mathrm{B}$ is the Boltzmann constant, $T_\mathrm{amp}$ is the equivalent input noise of the cryogenic amplifier and $T_\mathrm{eff}$ is the effective Johnson noise temperature considering the effect of quantum noise at the excitation frequency and the base temperature ($k_\mathrm{B}T_\mathrm{eff}=\hbar\omega \coth(\hbar \omega /k_\mathrm{B}T_\mathrm{base})$). At 4 K, the thermal Johnson noise is close to the noise associated with the first stage semiconductor microwave amplifier (5 K), requiring only very small parametric gain in the SQUID for the amplified noise to exceed the noise from the semiconductor amplifier. However, we expect that at lower temperature a stronger excitation can be applied to the SQUID to provide higher parametric gain in order to reduce the effect of amplifier noise before noise amplification becomes dominant. 
Fig.~\ref{fig:estimation}(a,b) shows the noise performance estimated from the data in Fig.~\ref{fig:raw_data}(b,c) using the above relations at 4 K and 100 mK. In the linear branch (right branch) the SQUID does not show enhancement in flux sensitivity. This also coincides with zero parametric gain and hence no noise rise. At 100 mK, biasing the SQUID at the nonlinear branch reduces the overall noise temperature of the amplification chain, therefore the noise performance is enhanced. In this branch, the complication arises that the reduction of the transduction coefficient at higher parametric gain eventually degrades the noise performance of the device, in particular at 4 K [see inset in Fig.~\ref{fig:estimation}(c)]. This effect is attributed to phase sensitive amplification in parametric amplifiers\cite{Levenson} where noise is amplified more than the transduction because the phase of the flux signal does not coincide with the amplified quadrature. 

To verify the noise model, we directly measured the noise at different points
on both linear and non-linear branches [Fig.~\ref{fig:estimation}(c)]. At each
different bias point we measured the spectral power density of the mixer
output. The flux noise was estimated by dividing the voltage spectral density
by the transduction coefficient computed from the area under the excitation peak.

At 4 K the best operating points lie on the linear branch where parametric gain is negligible. We obtained an effective flux noise of $80\;\mathrm{n}\Phi_0\mathrm{Hz}^{-1/2}$. This agrees well with an estimated best noise performance of $\sim 76\;\mathrm{n}\Phi_0\mathrm{Hz}^{-1/2}$ [Fig.~\ref{fig:estimation}(a)]. Nevertheless, the estimate in Fig.~\ref{fig:estimation}(b) predicts that at a base temperature below $\sim$300 mK where quantum noise dominates, one can harness the parametric gain in the nonlinear branch and reduce the flux noise to $30\;\mathrm{n}\Phi_0\mathrm{Hz}^{-1/2}$. According to Tesche and Clarke\cite{Tesche1977}, a DC SQUID gradiometer in the limit of $\beta \approx 1$ with equal inductance and junction critical current density would perform at a flux noise of $S^{1/2}_\Phi=8k_\mathrm{B}TL\sqrt{\pi LC}\approx 292\;\mathrm{n}\Phi_0 \mathrm{Hz}^{-1/2}$, ultimately quantum limited by $S^{1/2}_\Phi=hL=79\;\mathrm{n}\Phi_0 \mathrm{Hz}^{-1/2}$ at dilution refrigerator temperatures. We attribute improved noise performance of our devices to the advantages of dispersive readout. 

Note that the bandwidth decreases as the parametric gain increases. The bandwidth was determined by sweeping the ac signal frequency at the field coil. For the linear branch it is about 200 MHz and independent of the applied power. This value lies close to an estimate of the bandwidth of a simple LC resonator with a line impedance of $50\;\Omega$ and a capacitance of 30 pF ($1/2 \pi Z_0C\approx$ 100 MHz). The bandwidth at the nonlinear branch is a function of the parametric gain and thus a function of power [inset of Fig.~\ref{fig:estimation}(c)]. The gain-bandwidth product is constant in parametric amplifiers\cite{Vijay}, corresponding to a lower bandwidth at higher parametric gain. At 100 mK, the line noise at 6.6 GHz is quantum limited and a minimum parametric gain of 5 is needed to overcome the semiconductor amplifier noise. This value of gain corresponds to a bandwidth of approximately 20 MHz.\\

In conclusion, we have characterized a Nb scanning SQUID at 4 K. Using homodyne to readout our dispersive RF SQUIDs, we reached a flux noise of $S^{1/2}_\Phi = 80\;\mathrm{n}\Phi_0\mathrm{Hz}^{-1/2}$ with 200 MHz bandwidth at 4 K. Furthermore, estimations predict that at lower temperature and higher excitation power the flux noise decreases to $30\;\mathrm{n}\Phi_0\mathrm{Hz}^{-1/2}$ at a bandwidth of 20 MHz. Thus, our micro-SQUID shows flux sensitivity on par with the best reported micro- and nano-SQUIDs, sub-micron pickup loops and integrated on-chip field coils and a scanning microscopy compatible geometry. A reduced back action and dissipation compared to conventional DC SQUIDs can be expected because of the lower operating frequency and easier variation of the excitation power. Furthermore, heating effects commonly found in on-chip resistors can be avoided.\cite{Wellstood1994} These features make it ideal for single electron spin detection and manipulation with the goal of observing quantum coherent phenomena in mesoscopic samples.\\

This work was supported by the Alfried Krupp von Bohlen und Halbach Foundation and DFG grant BL 1197/3-1. The authors thank D. Yohannes, O. Mukhanov, M. Radparvar, and A. Kirichenko from \textit{Hypres Inc.} The SQUIDs used in this study were developed under an NSF IMR-MIP contract, Award number 0957616.
\bibliography{main}

\providecommand{\noopsort}[1]{}\providecommand{\singleletter}[1]{#1}%
\begin{thebibliography}{20}%
\makeatletter
\providecommand \@ifxundefined [1]{%
 \@ifx{#1\undefined}
}%
\providecommand \@ifnum [1]{%
 \ifnum #1\expandafter \@firstoftwo
 \else \expandafter \@secondoftwo
 \fi
}%
\providecommand \@ifx [1]{%
 \ifx #1\expandafter \@firstoftwo
 \else \expandafter \@secondoftwo
 \fi
}%
\providecommand \natexlab [1]{#1}%
\providecommand \enquote  [1]{``#1''}%
\providecommand \bibnamefont  [1]{#1}%
\providecommand \bibfnamefont [1]{#1}%
\providecommand \citenamefont [1]{#1}%
\providecommand \href@noop [0]{\@secondoftwo}%
\providecommand \href [0]{\begingroup \@sanitize@url \@href}%
\providecommand \@href[1]{\@@startlink{#1}\@@href}%
\providecommand \@@href[1]{\endgroup#1\@@endlink}%
\providecommand \@sanitize@url [0]{\catcode `\\12\catcode `\$12\catcode
  `\&12\catcode `\#12\catcode `\^12\catcode `\_12\catcode `\%12\relax}%
\providecommand \@@startlink[1]{}%
\providecommand \@@endlink[0]{}%
\providecommand \url  [0]{\begingroup\@sanitize@url \@url }%
\providecommand \@url [1]{\endgroup\@href {#1}{\urlprefix }}%
\providecommand \urlprefix  [0]{URL }%
\providecommand \Eprint [0]{\href }%
\providecommand \doibase [0]{http://dx.doi.org/}%
\providecommand \selectlanguage [0]{\@gobble}%
\providecommand \bibinfo  [0]{\@secondoftwo}%
\providecommand \bibfield  [0]{\@secondoftwo}%
\providecommand \translation [1]{[#1]}%
\providecommand \BibitemOpen [0]{}%
\providecommand \bibitemStop [0]{}%
\providecommand \bibitemNoStop [0]{.\EOS\space}%
\providecommand \EOS [0]{\spacefactor3000\relax}%
\providecommand \BibitemShut  [1]{\csname bibitem#1\endcsname}%
\let\auto@bib@innerbib\@empty
\bibitem [{\citenamefont {Martin}\ and\ \citenamefont
  {Wickramasinghe}(1987)}]{MFM}%
  \BibitemOpen
  \bibfield  {author} {\bibinfo {author} {\bibfnamefont {Y.}~\bibnamefont
  {Martin}}\ and\ \bibinfo {author} {\bibfnamefont {H.~K.}\ \bibnamefont
  {Wickramasinghe}},\ }\href {\doibase http://dx.doi.org/10.1063/1.97800}
  {\bibfield  {journal} {\bibinfo  {journal} {Applied Physics Letters}\
  }\textbf {\bibinfo {volume} {50}},\ \bibinfo {pages} {1455} (\bibinfo {year}
  {1987})}\BibitemShut {NoStop}%
\bibitem [{\citenamefont {Chang}\ \emph {et~al.}(1992)\citenamefont {Chang},
  \citenamefont {Hallen}, \citenamefont {Harriott}, \citenamefont {Hess},
  \citenamefont {Kao}, \citenamefont {Kwo}, \citenamefont {Miller},
  \citenamefont {Wolfe}, \citenamefont {van~der Ziel},\ and\ \citenamefont
  {Chang}}]{Hall}%
  \BibitemOpen
  \bibfield  {author} {\bibinfo {author} {\bibfnamefont {A.~M.}\ \bibnamefont
  {Chang}}, \bibinfo {author} {\bibfnamefont {H.~D.}\ \bibnamefont {Hallen}},
  \bibinfo {author} {\bibfnamefont {L.}~\bibnamefont {Harriott}}, \bibinfo
  {author} {\bibfnamefont {H.~F.}\ \bibnamefont {Hess}}, \bibinfo {author}
  {\bibfnamefont {H.~L.}\ \bibnamefont {Kao}}, \bibinfo {author} {\bibfnamefont
  {J.}~\bibnamefont {Kwo}}, \bibinfo {author} {\bibfnamefont {R.~E.}\
  \bibnamefont {Miller}}, \bibinfo {author} {\bibfnamefont {R.}~\bibnamefont
  {Wolfe}}, \bibinfo {author} {\bibfnamefont {J.}~\bibnamefont {van~der Ziel}},
  \ and\ \bibinfo {author} {\bibfnamefont {T.~Y.}\ \bibnamefont {Chang}},\
  }\href {\doibase http://dx.doi.org/10.1063/1.108334} {\bibfield  {journal}
  {\bibinfo  {journal} {Applied Physics Letters}\ }\textbf {\bibinfo {volume}
  {61}},\ \bibinfo {pages} {1974} (\bibinfo {year} {1992})}\BibitemShut
  {NoStop}%
\bibitem [{\citenamefont {Kirtley}\ and\ \citenamefont
  {Wikswo}(1999)}]{John_review}%
  \BibitemOpen
  \bibfield  {author} {\bibinfo {author} {\bibfnamefont {J.}~\bibnamefont
  {Kirtley}}\ and\ \bibinfo {author} {\bibfnamefont {J.~P.}\ \bibnamefont
  {Wikswo}},\ }\href@noop {} {\bibfield  {journal} {\bibinfo  {journal} {Annu.
  Rev. Mater. Sci}\ }\textbf {\bibinfo {volume} {29}},\ \bibinfo {pages} {117}
  (\bibinfo {year} {1999})}\BibitemShut {NoStop}%
\bibitem [{\citenamefont {Hasselbach}, \citenamefont {Veauvy},\ and\
  \citenamefont {Mailly}(2000)}]{Hasselbach}%
  \BibitemOpen
  \bibfield  {author} {\bibinfo {author} {\bibfnamefont {K.}~\bibnamefont
  {Hasselbach}}, \bibinfo {author} {\bibfnamefont {C.}~\bibnamefont {Veauvy}},
  \ and\ \bibinfo {author} {\bibfnamefont {D.}~\bibnamefont {Mailly}},\
  }\href@noop {} {\bibfield  {journal} {\bibinfo  {journal} {Physica C:
  Superconductivity}\ }\textbf {\bibinfo {volume} {332}},\ \bibinfo {pages}
  {140 } (\bibinfo {year} {2000})}\BibitemShut {NoStop}%
\bibitem [{\citenamefont {Maletinsky}\ \emph {et~al.}(2012)\citenamefont
  {Maletinsky}, \citenamefont {Hong}, \citenamefont {Grinolds}, \citenamefont
  {Hausmann}, \citenamefont {Lukin}, \citenamefont {Walsworth}, \citenamefont
  {Loncar},\ and\ \citenamefont {Yacoby}}]{NV}%
  \BibitemOpen
  \bibfield  {author} {\bibinfo {author} {\bibfnamefont {P.}~\bibnamefont
  {Maletinsky}}, \bibinfo {author} {\bibfnamefont {S.}~\bibnamefont {Hong}},
  \bibinfo {author} {\bibfnamefont {M.~S.}\ \bibnamefont {Grinolds}}, \bibinfo
  {author} {\bibfnamefont {B.}~\bibnamefont {Hausmann}}, \bibinfo {author}
  {\bibfnamefont {M.~D.}\ \bibnamefont {Lukin}}, \bibinfo {author}
  {\bibfnamefont {R.~L.}\ \bibnamefont {Walsworth}}, \bibinfo {author}
  {\bibfnamefont {M.}~\bibnamefont {Loncar}}, \ and\ \bibinfo {author}
  {\bibfnamefont {A.}~\bibnamefont {Yacoby}},\ }\href@noop {} {\bibfield
  {journal} {\bibinfo  {journal} {Nature Nanotechnology}\ }\textbf {\bibinfo
  {volume} {7}},\ \bibinfo {pages} {324} (\bibinfo {year} {2012})}\BibitemShut
  {NoStop}%
\bibitem [{\citenamefont {Tsuei}\ \emph {et~al.}(1994)\citenamefont {Tsuei},
  \citenamefont {Kirtley}, \citenamefont {Chi}, \citenamefont {Yu-Jahnes},
  \citenamefont {Gupta}, \citenamefont {Shaw}, \citenamefont {Sun},\ and\
  \citenamefont {Ketchen}}]{John_pairing}%
  \BibitemOpen
  \bibfield  {author} {\bibinfo {author} {\bibfnamefont {C.~C.}\ \bibnamefont
  {Tsuei}}, \bibinfo {author} {\bibfnamefont {J.~R.}\ \bibnamefont {Kirtley}},
  \bibinfo {author} {\bibfnamefont {C.~C.}\ \bibnamefont {Chi}}, \bibinfo
  {author} {\bibfnamefont {L.~S.}\ \bibnamefont {Yu-Jahnes}}, \bibinfo {author}
  {\bibfnamefont {A.}~\bibnamefont {Gupta}}, \bibinfo {author} {\bibfnamefont
  {T.}~\bibnamefont {Shaw}}, \bibinfo {author} {\bibfnamefont {J.~Z.}\
  \bibnamefont {Sun}}, \ and\ \bibinfo {author} {\bibfnamefont {M.~B.}\
  \bibnamefont {Ketchen}},\ }\href {\doibase 10.1103/PhysRevLett.73.593}
  {\bibfield  {journal} {\bibinfo  {journal} {Phys. Rev. Lett.}\ }\textbf
  {\bibinfo {volume} {73}},\ \bibinfo {pages} {593} (\bibinfo {year}
  {1994})}\BibitemShut {NoStop}%
\bibitem [{\citenamefont {Bluhm}\ \emph {et~al.}(2009)\citenamefont {Bluhm},
  \citenamefont {Koshnick}, \citenamefont {Bert}, \citenamefont {Huber},\ and\
  \citenamefont {Moler}}]{Hendrik_persistent}%
  \BibitemOpen
  \bibfield  {author} {\bibinfo {author} {\bibfnamefont {H.}~\bibnamefont
  {Bluhm}}, \bibinfo {author} {\bibfnamefont {N.~C.}\ \bibnamefont {Koshnick}},
  \bibinfo {author} {\bibfnamefont {J.~A.}\ \bibnamefont {Bert}}, \bibinfo
  {author} {\bibfnamefont {M.~E.}\ \bibnamefont {Huber}}, \ and\ \bibinfo
  {author} {\bibfnamefont {K.~A.}\ \bibnamefont {Moler}},\ }\href {\doibase
  10.1103/PhysRevLett.102.136802} {\bibfield  {journal} {\bibinfo  {journal}
  {Phys. Rev. Lett.}\ }\textbf {\bibinfo {volume} {102}},\ \bibinfo {pages}
  {136802} (\bibinfo {year} {2009})}\BibitemShut {NoStop}%
\bibitem [{\citenamefont {Nowack}\ \emph {et~al.}(2013)\citenamefont {Nowack},
  \citenamefont {Spanton}, \citenamefont {Baenninger}, \citenamefont {Koenig},
  \citenamefont {Kirtley}, \citenamefont {Kalisky}, \citenamefont {Ames},
  \citenamefont {Leubner}, \citenamefont {Bruene}, \citenamefont {Buhmann},
  \citenamefont {Molenkamp}, \citenamefont {Goldhaber-Gordon},\ and\
  \citenamefont {Moler}}]{katja}%
  \BibitemOpen
  \bibfield  {author} {\bibinfo {author} {\bibfnamefont {K.~C.}\ \bibnamefont
  {Nowack}}, \bibinfo {author} {\bibfnamefont {E.~M.}\ \bibnamefont {Spanton}},
  \bibinfo {author} {\bibfnamefont {M.}~\bibnamefont {Baenninger}}, \bibinfo
  {author} {\bibfnamefont {M.}~\bibnamefont {Koenig}}, \bibinfo {author}
  {\bibfnamefont {J.~R.}\ \bibnamefont {Kirtley}}, \bibinfo {author}
  {\bibfnamefont {B.}~\bibnamefont {Kalisky}}, \bibinfo {author} {\bibfnamefont
  {C.}~\bibnamefont {Ames}}, \bibinfo {author} {\bibfnamefont {P.}~\bibnamefont
  {Leubner}}, \bibinfo {author} {\bibfnamefont {C.}~\bibnamefont {Bruene}},
  \bibinfo {author} {\bibfnamefont {H.}~\bibnamefont {Buhmann}}, \bibinfo
  {author} {\bibfnamefont {L.~W.}\ \bibnamefont {Molenkamp}}, \bibinfo {author}
  {\bibfnamefont {D.}~\bibnamefont {Goldhaber-Gordon}}, \ and\ \bibinfo
  {author} {\bibfnamefont {K.~A.}\ \bibnamefont {Moler}},\ }\href@noop {}
  {\bibfield  {journal} {\bibinfo  {journal} {Nat. Mater.}\ }\textbf {\bibinfo
  {volume} {12}},\ \bibinfo {pages} {787 } (\bibinfo {year}
  {2013})}\BibitemShut {NoStop}%
\bibitem [{\citenamefont {Huber}\ \emph {et~al.}(2008)\citenamefont {Huber},
  \citenamefont {Koshnick}, \citenamefont {Bluhm}, \citenamefont {Archuleta},
  \citenamefont {Azua}, \citenamefont {Bjoernsson}, \citenamefont {Gardner},
  \citenamefont {Halloran}, \citenamefont {Lucero},\ and\ \citenamefont
  {Moler}}]{Huber_SQUID}%
  \BibitemOpen
  \bibfield  {author} {\bibinfo {author} {\bibfnamefont {M.~E.}\ \bibnamefont
  {Huber}}, \bibinfo {author} {\bibfnamefont {N.~C.}\ \bibnamefont {Koshnick}},
  \bibinfo {author} {\bibfnamefont {H.}~\bibnamefont {Bluhm}}, \bibinfo
  {author} {\bibfnamefont {L.~J.}\ \bibnamefont {Archuleta}}, \bibinfo {author}
  {\bibfnamefont {T.}~\bibnamefont {Azua}}, \bibinfo {author} {\bibfnamefont
  {P.~G.}\ \bibnamefont {Bjoernsson}}, \bibinfo {author} {\bibfnamefont
  {B.~W.}\ \bibnamefont {Gardner}}, \bibinfo {author} {\bibfnamefont {S.~T.}\
  \bibnamefont {Halloran}}, \bibinfo {author} {\bibfnamefont {E.~A.}\
  \bibnamefont {Lucero}}, \ and\ \bibinfo {author} {\bibfnamefont {K.~A.}\
  \bibnamefont {Moler}},\ }\href {\doibase 10.1063/1.2932341} {\bibfield
  {journal} {\bibinfo  {journal} {Rev. Sci. Instrum.}\ }\textbf {\bibinfo
  {volume} {79}} (\bibinfo {year} {2008}),\ 10.1063/1.2932341}\BibitemShut
  {NoStop}%
\bibitem [{\citenamefont {Vasyukov}\ \emph {et~al.}(2013)\citenamefont
  {Vasyukov}, \citenamefont {Anahory}, \citenamefont {Embon}, \citenamefont
  {Halbertal}, \citenamefont {Cuppens}, \citenamefont {Neeman}, \citenamefont
  {Finkler}, \citenamefont {Segev}, \citenamefont {Myasoedov}, \citenamefont
  {Rappaport}, \citenamefont {Huber},\ and\ \citenamefont {Zeldov}}]{Zeldov}%
  \BibitemOpen
  \bibfield  {author} {\bibinfo {author} {\bibfnamefont {D.}~\bibnamefont
  {Vasyukov}}, \bibinfo {author} {\bibfnamefont {Y.}~\bibnamefont {Anahory}},
  \bibinfo {author} {\bibfnamefont {L.}~\bibnamefont {Embon}}, \bibinfo
  {author} {\bibfnamefont {D.}~\bibnamefont {Halbertal}}, \bibinfo {author}
  {\bibfnamefont {J.}~\bibnamefont {Cuppens}}, \bibinfo {author} {\bibfnamefont
  {L.}~\bibnamefont {Neeman}}, \bibinfo {author} {\bibfnamefont
  {A.}~\bibnamefont {Finkler}}, \bibinfo {author} {\bibfnamefont
  {Y.}~\bibnamefont {Segev}}, \bibinfo {author} {\bibfnamefont
  {Y.}~\bibnamefont {Myasoedov}}, \bibinfo {author} {\bibfnamefont {M.~L.}\
  \bibnamefont {Rappaport}}, \bibinfo {author} {\bibfnamefont {M.~E.}\
  \bibnamefont {Huber}}, \ and\ \bibinfo {author} {\bibfnamefont
  {E.}~\bibnamefont {Zeldov}},\ }\href
  {http://dx.doi.org/10.1038/nnano.2013.169} {\bibfield  {journal} {\bibinfo
  {journal} {Nat Nano}\ }\textbf {\bibinfo {volume} {8}},\ \bibinfo {pages}
  {639} (\bibinfo {year} {2013})},\ \bibinfo {note} {letter}\BibitemShut
  {NoStop}%
\bibitem [{\citenamefont {Hatridge}\ \emph {et~al.}(2011)\citenamefont
  {Hatridge}, \citenamefont {Vijay}, \citenamefont {Slichter}, \citenamefont
  {Clarke},\ and\ \citenamefont {Siddiqi}}]{Vijay}%
  \BibitemOpen
  \bibfield  {author} {\bibinfo {author} {\bibfnamefont {M.}~\bibnamefont
  {Hatridge}}, \bibinfo {author} {\bibfnamefont {R.}~\bibnamefont {Vijay}},
  \bibinfo {author} {\bibfnamefont {D.~H.}\ \bibnamefont {Slichter}}, \bibinfo
  {author} {\bibfnamefont {J.}~\bibnamefont {Clarke}}, \ and\ \bibinfo {author}
  {\bibfnamefont {I.}~\bibnamefont {Siddiqi}},\ }\href {\doibase
  10.1103/PhysRevB.83.134501} {\bibfield  {journal} {\bibinfo  {journal} {Phys.
  Rev. B}\ }\textbf {\bibinfo {volume} {83}},\ \bibinfo {pages} {134501}
  (\bibinfo {year} {2011})}\BibitemShut {NoStop}%
\bibitem [{\citenamefont {Levenson-Falk}\ \emph {et~al.}(2013)\citenamefont
  {Levenson-Falk}, \citenamefont {Vijay}, \citenamefont {Antler},\ and\
  \citenamefont {Siddiqi}}]{Levenson}%
  \BibitemOpen
  \bibfield  {author} {\bibinfo {author} {\bibfnamefont {E.~M.}\ \bibnamefont
  {Levenson-Falk}}, \bibinfo {author} {\bibfnamefont {R.}~\bibnamefont
  {Vijay}}, \bibinfo {author} {\bibfnamefont {N.}~\bibnamefont {Antler}}, \
  and\ \bibinfo {author} {\bibfnamefont {I.}~\bibnamefont {Siddiqi}},\ }\href
  {http://stacks.iop.org/0953-2048/26/i=5/a=055015} {\bibfield  {journal}
  {\bibinfo  {journal} {Superconductor Science and Technology}\ }\textbf
  {\bibinfo {volume} {26}},\ \bibinfo {pages} {055015} (\bibinfo {year}
  {2013})}\BibitemShut {NoStop}%
\bibitem [{\citenamefont {Castellanos-Beltran}\ \emph
  {et~al.}(2008)\citenamefont {Castellanos-Beltran}, \citenamefont {Irwin},
  \citenamefont {Hilton}, \citenamefont {Vale},\ and\ \citenamefont
  {W.}}]{Castellanos_nature}%
  \BibitemOpen
  \bibfield  {author} {\bibinfo {author} {\bibfnamefont {M.~A.}\ \bibnamefont
  {Castellanos-Beltran}}, \bibinfo {author} {\bibfnamefont {K.~D.}\
  \bibnamefont {Irwin}}, \bibinfo {author} {\bibfnamefont {G.~C.}\ \bibnamefont
  {Hilton}}, \bibinfo {author} {\bibfnamefont {L.~R.}\ \bibnamefont {Vale}}, \
  and\ \bibinfo {author} {\bibfnamefont {L.~K.}\ \bibnamefont {W.}},\
  }\href@noop {} {\bibfield  {journal} {\bibinfo  {journal} {Nature Physics}\
  }\textbf {\bibinfo {volume} {4}},\ \bibinfo {pages} {931} (\bibinfo {year}
  {2008})}\BibitemShut {NoStop}%
\bibitem [{\citenamefont {Vijay}, \citenamefont {Devoret},\ and\ \citenamefont
  {Siddiqi}(2009)}]{Sidiqi_review}%
  \BibitemOpen
  \bibfield  {author} {\bibinfo {author} {\bibfnamefont {R.}~\bibnamefont
  {Vijay}}, \bibinfo {author} {\bibfnamefont {M.~H.}\ \bibnamefont {Devoret}},
  \ and\ \bibinfo {author} {\bibfnamefont {I.}~\bibnamefont {Siddiqi}},\ }\href
  {\doibase http://dx.doi.org/10.1063/1.3224703} {\bibfield  {journal}
  {\bibinfo  {journal} {Review of Scientific Instruments}\ }\textbf {\bibinfo
  {volume} {80}},\ \bibinfo {eid} {111101} (\bibinfo {year}
  {2009})}\BibitemShut {NoStop}%
\bibitem [{\citenamefont {O'Brien}\ \emph {et~al.}(2014)\citenamefont
  {O'Brien}, \citenamefont {Macklin}, \citenamefont {Siddiqi},\ and\
  \citenamefont {Zhang}}]{Sidiqi_HBW}%
  \BibitemOpen
  \bibfield  {author} {\bibinfo {author} {\bibfnamefont {K.}~\bibnamefont
  {O'Brien}}, \bibinfo {author} {\bibfnamefont {C.}~\bibnamefont {Macklin}},
  \bibinfo {author} {\bibfnamefont {I.}~\bibnamefont {Siddiqi}}, \ and\
  \bibinfo {author} {\bibfnamefont {X.}~\bibnamefont {Zhang}},\ }\href
  {\doibase 10.1103/PhysRevLett.113.157001} {\bibfield  {journal} {\bibinfo
  {journal} {Phys. Rev. Lett.}\ }\textbf {\bibinfo {volume} {113}},\ \bibinfo
  {pages} {157001} (\bibinfo {year} {2014})}\BibitemShut {NoStop}%
\bibitem [{\citenamefont {Koshnick}\ \emph {et~al.}(2008)\citenamefont
  {Koshnick}, \citenamefont {Huber}, \citenamefont {Bert}, \citenamefont
  {Hicks}, \citenamefont {Large}, \citenamefont {Edwards},\ and\ \citenamefont
  {Moler}}]{koshnik}%
  \BibitemOpen
  \bibfield  {author} {\bibinfo {author} {\bibfnamefont {N.~C.}\ \bibnamefont
  {Koshnick}}, \bibinfo {author} {\bibfnamefont {M.~E.}\ \bibnamefont {Huber}},
  \bibinfo {author} {\bibfnamefont {J.~A.}\ \bibnamefont {Bert}}, \bibinfo
  {author} {\bibfnamefont {C.~W.}\ \bibnamefont {Hicks}}, \bibinfo {author}
  {\bibfnamefont {J.}~\bibnamefont {Large}}, \bibinfo {author} {\bibfnamefont
  {H.}~\bibnamefont {Edwards}}, \ and\ \bibinfo {author} {\bibfnamefont
  {K.~A.}\ \bibnamefont {Moler}},\ }\href {\doibase
  http://dx.doi.org/10.1063/1.3046098} {\bibfield  {journal} {\bibinfo
  {journal} {Applied Physics Letters}\ }\textbf {\bibinfo {volume} {93}},\
  \bibinfo {eid} {243101} (\bibinfo {year} {2008}),\
  http://dx.doi.org/10.1063/1.3046098}\BibitemShut {NoStop}%
\bibitem [{\citenamefont {Kirtley}\ \emph {et~al.}(2016)\citenamefont
  {Kirtley}, \citenamefont {Paulius}, \citenamefont {Rosenberg}, \citenamefont
  {Palmstrom}, \citenamefont {Holland}, \citenamefont {Spanton}, \citenamefont
  {Schiessl}, \citenamefont {Jermain}, \citenamefont {Gibbons}, \citenamefont
  {Fung}, \citenamefont {Huber}, \citenamefont {Ralph}, \citenamefont
  {Ketchen}, \citenamefont {GibsonJr.},\ and\ \citenamefont
  {Moler}}]{Kirtley2016}%
  \BibitemOpen
  \bibfield  {author} {\bibinfo {author} {\bibfnamefont {J.~R.}\ \bibnamefont
  {Kirtley}}, \bibinfo {author} {\bibfnamefont {L.}~\bibnamefont {Paulius}},
  \bibinfo {author} {\bibfnamefont {A.~J.}\ \bibnamefont {Rosenberg}}, \bibinfo
  {author} {\bibfnamefont {J.~C.}\ \bibnamefont {Palmstrom}}, \bibinfo {author}
  {\bibfnamefont {C.~M.}\ \bibnamefont {Holland}}, \bibinfo {author}
  {\bibfnamefont {E.~M.}\ \bibnamefont {Spanton}}, \bibinfo {author}
  {\bibfnamefont {D.}~\bibnamefont {Schiessl}}, \bibinfo {author}
  {\bibfnamefont {C.~L.}\ \bibnamefont {Jermain}}, \bibinfo {author}
  {\bibfnamefont {J.}~\bibnamefont {Gibbons}}, \bibinfo {author} {\bibfnamefont
  {Y.-K.-K.}\ \bibnamefont {Fung}}, \bibinfo {author} {\bibfnamefont {M.~E.}\
  \bibnamefont {Huber}}, \bibinfo {author} {\bibfnamefont {D.~C.}\ \bibnamefont
  {Ralph}}, \bibinfo {author} {\bibfnamefont {M.~B.}\ \bibnamefont {Ketchen}},
  \bibinfo {author} {\bibfnamefont {G.~W.}\ \bibnamefont {GibsonJr.}}, \ and\
  \bibinfo {author} {\bibfnamefont {K.~A.}\ \bibnamefont {Moler}},\ }\href
  {\doibase 10.1063/1.4961982} {\bibfield  {journal} {\bibinfo  {journal}
  {Review of Scientific Instruments}\ }\textbf {\bibinfo {volume} {87}},\
  \bibinfo {pages} {093702} (\bibinfo {year} {2016})},\ \Eprint
  {http://arxiv.org/abs/https://doi.org/10.1063/1.4961982}
  {https://doi.org/10.1063/1.4961982} \BibitemShut {NoStop}%
\bibitem [{\citenamefont {Clarke}\ and\ \citenamefont
  {Braginski}(2003)}]{handbook}%
  \BibitemOpen
  \bibinfo {editor} {\bibfnamefont {J.}~\bibnamefont {Clarke}}\ and\ \bibinfo
  {editor} {\bibfnamefont {A.~J.}\ \bibnamefont {Braginski}},\ eds.,\
  \href@noop {} {\emph {\bibinfo {title} {The SQUID Handbook}}}\ (\bibinfo
  {publisher} {Wiley-VCH},\ \bibinfo {year} {2003})\BibitemShut {NoStop}%
\bibitem [{\citenamefont {Tesche}\ and\ \citenamefont
  {Clarke}(1977)}]{Tesche1977}%
  \BibitemOpen
  \bibfield  {author} {\bibinfo {author} {\bibfnamefont {C.~D.}\ \bibnamefont
  {Tesche}}\ and\ \bibinfo {author} {\bibfnamefont {J.}~\bibnamefont
  {Clarke}},\ }\href {\doibase 10.1007/BF00655097} {\bibfield  {journal}
  {\bibinfo  {journal} {Journal of Low Temperature Physics}\ }\textbf {\bibinfo
  {volume} {29}},\ \bibinfo {pages} {301} (\bibinfo {year} {1977})}\BibitemShut
  {NoStop}%
\bibitem [{\citenamefont {Wellstood}, \citenamefont {Urbina},\ and\
  \citenamefont {Clarke}(1994)}]{Wellstood1994}%
  \BibitemOpen
  \bibfield  {author} {\bibinfo {author} {\bibfnamefont {F.~C.}\ \bibnamefont
  {Wellstood}}, \bibinfo {author} {\bibfnamefont {C.}~\bibnamefont {Urbina}}, \
  and\ \bibinfo {author} {\bibfnamefont {J.}~\bibnamefont {Clarke}},\ }\href
  {\doibase 10.1103/PhysRevB.49.5942} {\bibfield  {journal} {\bibinfo
  {journal} {Phys. Rev. B}\ }\textbf {\bibinfo {volume} {49}},\ \bibinfo
  {pages} {5942} (\bibinfo {year} {1994})}\BibitemShut {NoStop}%
\end{thebibliography}%
\end{document}